\documentclass[aps,prl,twocolumn,amsmath,amssymb,superscriptaddress,citeautoscript,longbibliography,floatfix]{revtex4-1}
\newcommand{\angstrom}{\textup{\AA}}
\usepackage[allow-number-unit-breaks]{siunitx}
\usepackage{graphicx}
\usepackage{dcolumn}
\usepackage{bm}
\usepackage{color}
\usepackage{bbding}
\usepackage{pifont}

\usepackage{mathrsfs}
\usepackage{amsfonts}
\usepackage{multirow}
\usepackage{natbib}
\usepackage{epsf}
\usepackage{epsfig}
\usepackage{epstopdf}
\usepackage{amsmath}
\usepackage{hyperref}
\hypersetup{
    colorlinks=true,
    linkcolor=blue,
    filecolor=cyan,      
    citecolor=blue,
    urlcolor=blue,
    }

\begin{document}

\title{Manipulation of Ferroelectric Polarization Under Near-Gap Light}

\author{Hanli Cui}
\affiliation{Center for Alloy Innovation and Design, State Key Laboratory for Mechanical Behavior of Materials, Xi'an Jiaotong University, Xi'an, 710049, China}

\author{Jian Zhou}\email{jianzhou@xjtu.edu.cn}
\affiliation{Center for Alloy Innovation and Design, State Key Laboratory for Mechanical Behavior of Materials, Xi'an Jiaotong University, Xi'an, 710049, China}

\begin{abstract}
Noncontacting light control of ferroelectric (FE) order has been receiving extensive interest in condensed matter physics. In this work, we investigate near-resonant light irradiation that induces structure-dependent photoexcited carrier density along the transition pathway. We combine a minimal Rice-Mele model containing intersite density-density correlations with \textit{ab initio} calculations for realistic materials such as $\rm GeTe$ and $\rm SnTe$. If the photon energy is below-gap without carrier excitation, only off-resonant scattering occurs and the FE order tends to be diminished. Slightly increasing photon energy at the near-gap regime would lead to a centrosymmetric non-polar (NP) to FE phase transition, which is not thermally accessible in conventional pressure-temperature phase diagram. This is in contrast to the well-above-gap light situation that tends to suppress the FE order under a fixed amount of photoexcitation. We apply phenomenological Landau theory and show that they belong to the continuous and first-order phase transitions, respectively. The occupation constrained calculations suggest that these reversible NP--FE phase transitions could be triggered under moderate light fluence. Our mechanism provides a potential route to reversible light-stabilized polar phases with reduced lattice heating and marginal lattice damages.
\end{abstract}

\maketitle
\textit{Introduction.}---Efficient and ultrafast control of ferroic orders is one of the main topics in condensed matter physics, material science, and information technology. Among various ferroic orders in quantum materials, ferroelectric (FE) systems have garnered significant attention due to their nonvolatile nature and multiple applications \cite{Yonemitsu06JPSJ.75.011008,Nasu_2001JPhys201,Mizouchi01PT082001,Chandra_2017aa82d2,Ahn17PhysRevLett.119.057601,Kundys12PhysRevB.85.092301}. While electrically-induced FE phase transition has been widely adopted, noncontact optical manipulation of FE orders has become one of the attractive fields \cite{Mehboudi2016,Marini21PhysRevLett.127.257401,Paillard19PhysRevLett.123.087601,Peng24PhysRevLett.132.116601,Peng20sciadv.abd1618,Chen20PhysRevB.102.184115,Zhou23PhysRevApplied.20.024020,Yang24PhysRevLett.133.136902,Wang24PhysRevLett.133.126801,Gao24PhysRevLett.133.196801,Feng25PhysRevB.111.024305,Chen24PhysRevB.109.094102,mogi2025arixiv14800,Liu26PRLfv5l-rjyv}, as light usually does not introduce unwanted atomic impurities and can be well-tuned in experiments.

When the photon energy is well-above the bandgap ($E_g$) of semiconductors it irradiates, a certain amount of electron-hole (e-h) pairs are excited. The adiabatic potential energy landscape ($\Phi_Q$ with $Q$ denoting generalized coordinates) changes, yielding a photoinduced lattice deformation and/or a phase transition. According to previous works \cite{Paillard23PhysRevB.107.054107,Furci24PhysRevLett.132.236101,Huber24sciadv.adl4481,Rurali24PhysRevLett.133.116401,Chen18PhysRevLett.120.185701,Liu22pnas.2122534119,mogi2025arixiv14800,Dansou25PRMsbcn-pkv4,Haleoot2017}, well-above-gap light irradiating onto a FE material usually triggers a transformation toward its non-polar (NP) parental phase, namely, a FE-NP pathway. A similar example is the seminal Su-Schrieffer-Heeger model in which the dimerization could be suppressed \cite{Heeger88RevModPhys.60.781}. During this process, one assumes a fixed $n_{eh}$, corresponding to a canonical-like closed boundary condition with long carrier lifetime. This approximation is accurate when the carrier mobility and leakage to environment (as compared with transition kinetics) are sufficiently low, or as the light absorbance $A(\omega)$ barely changes during phase transition.

\begin{figure}[tb]
    \centering
    \includegraphics[width=0.48\textwidth]{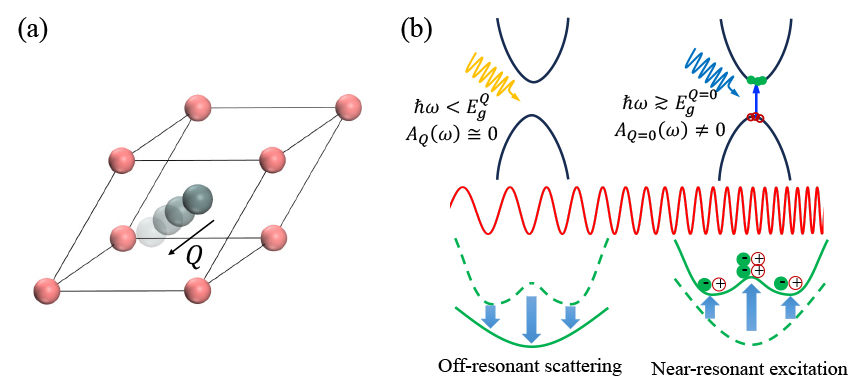}
    \caption{Near-gap light-induced phase transition. (a) Transition between the NP and FE as described by displacement mode $Q$, along which the band structure and optical absorbance vary significantly. For the group IV telluride compounds that exhibit (distorted) rocksalt phase, such $Q$ is along the $\langle 111\rangle$ directions. (b) Schematic plot of near-gap light effect on energy landscape $\Phi_Q$. Dashed green curves indicate the intact pattern (dark state), and the solid curves are light-dressed profile.}
    \label{fig:geo-ivvi}
\end{figure}

We note that when the incident photon energy is near the bandgap of reference phase (e.g., $\hbar\omega\simeq E_g^{Q=0}$), the significant variation of $E_g$ with $Q$ indicates that the light absorbance spectrum $A(\omega,Q)$ becomes strongly structure-dependent (Fig. \ref{fig:geo-ivvi}). This is different from the scenario of a fixed photoexcited charge carrier density $n_{eh}$ situation, and remains largely unexplored in prior works. Addressing this gap is beneficial for understanding how near-gap light controls lattice distortions and electronic behaviors, potentially enabling non-thermic reversible phase transition and discovery of hidden phases.

In this work, we perform theoretical analysis and numerical simulations to explore how near-gap light modulates $\Phi_Q$ and triggers non-thermic phase transitions. We adopt a minimal 1D Rice-Mele model \cite{Rice82PhysRevLett.49.1455} including intersite bond correlations. Two typical distinct frequencies are considered [Fig. \ref{fig:geo-ivvi}(b)]. When the photon energy is below the bandgap (off-resonant regime), light modulation on $\Phi_Q$ tends to stabilize the high-symmetric NP state, yet no photoexcitations emerge ($n_{eh}=0$). On the contrary, once the photon energy is enhanced near-resonant ($\hbar\omega\gtrsim E_g$), we consider structure-dependent absorbance $A(\omega,Q)$ effect, leading to a double well potential shape. Hence, a NP-FE phase transition could occur, so that one can expect an all-optical reversible phase transition cycle. Note that this is different from previous terahertz-induced FE emergence works that directly excite the infrared-active phonons \cite{Nova19science.aaw4911,Li19science.aaw4913,Shin22PhysRevLett.129.167401,Shin24PhysRevLett.132.016603}.

In order to further illustrate our prediction in realistic materials, we focus on group IV tellurides (specifically $\rm GeTe$ and $\rm SnTe$) as prototypical material examples using \textit{ab initio} calculations, in which the FE order could be optically switched on/off between cubic $Fm\bar{3}m$ ($Q=0$) and rhombohedral $R3m$ ($Q\neq 0$) phases. The direct and low gap feature ensures ultrafast carrier lifetime with marginal electron-phonon conversion, favoring adiabatic light-dressing condition and weak overheating requirement during reversible phase transition. Our first-principles density functional theory (DFT) calculations at both the ground and excited states suggest that transitions between $Fm\bar{3}m$ and $R3m$ could occur under moderate light fluences (a few to a few hundred $\mathrm{mJ/cm^2}$). In particular, the FE phase of $\mathrm{GeTe}$ could be driven to NP phase under off-resonant light. On the contrary, near-gap light is predicted to stabilize a polar distortion in NP tellurides such as $\rm SnTe$ and $\rm PbTe$, which is not thermally accessible in traditional phase diagram.

\begin{figure}[tb]
    \centering
    \includegraphics[width=0.48\textwidth]{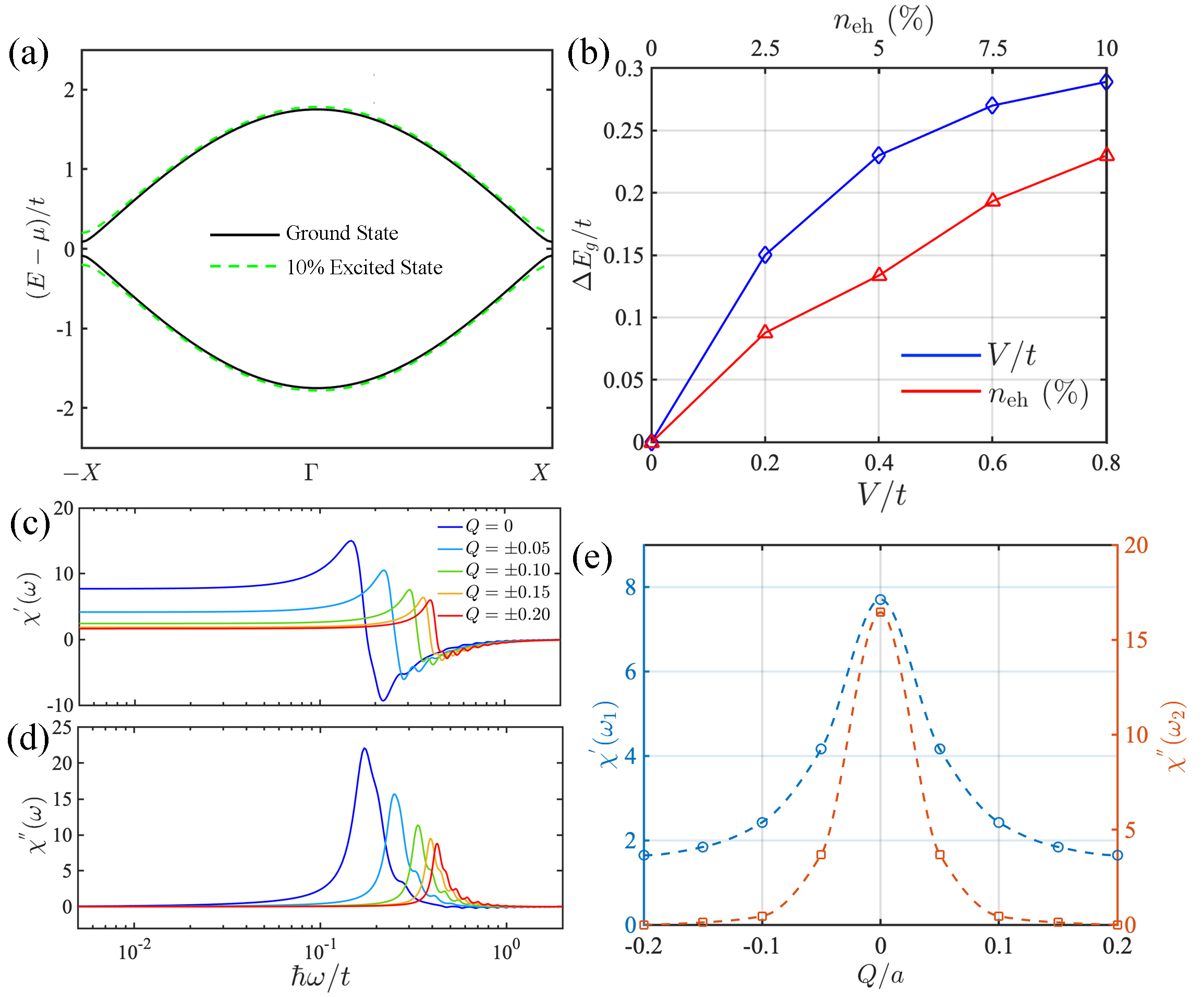}
    \caption{Minimal Rice-Mele model results. (a) Band dispersion of NP $Q=0$ state without (black solid) and with $n_{eh}=0.2\,e/\mathrm{f.u.}$ excitation (green dashed). $\mu$ indicates Fermi level at the half-filling. (b) The bandgap enhancement $\Delta E_g$ value with respect to parameter $V$ (blue diamond, keeping other parameters unchanged and a $n_{eh}=0.2\,e/\mathrm{f.u.}$ excitation) and e-h excitation concentration $n_{eh}$ (red triangle). (c) and (d) are frequency-dependent $\chi'(\omega)$ and $\chi''(\omega)$ under different $Q$. (e) Variation of $\chi'(\omega_1=0.1t)$ and $\chi''(\omega_2=0.2t)$ with $Q$ at the off-resonant regime and on-resonant regimes, respectively. One sees that they both exponentially reduce to lower values for large $|Q|$, indicating that near-gap light significantly affects the NP state.}
    \label{fig:tb}
\end{figure}

\textit{Tight binding model.}---We first adopt a simple and general model to conduct the analysis on the structure-dependent electronic behavior and optical feature. In order to describe FE phase, at least two different atoms should be included. Hence, the minimal model is the one-dimensional (1D) Rice-Mele model. This is used as a general model to analyze our mechanism, and the realistic material calculations are further guided by the results of this simple model and performed via DFT simulations. In a unit cell there are two sites $A$ and $B$. Under the basis set of $[|A_{\uparrow}\rangle,|B_{\uparrow}\rangle,|A_{\downarrow}\rangle,|B_{\downarrow}\rangle]^{\mathrm{T}}$, the single particle Hamiltonian reads
\begin{equation}\label{eq:Ham0}
    \begin{split}
        H_0=-2t\cos\frac{ak}{2}(\tau_x\cos kQ-\tau_y\sin kQ)\sigma_0+\Delta\tau_z\sigma_0,
    \end{split}
\end{equation}
where $\sigma$ and $\tau$ represent Pauli matrices for spin and sublattice, respectively. $a=1$ is the lattice constant, $t$ and $\Delta$ measure the hopping and on-site potential difference, respectively. Finite $\Delta$ indicates charge transfer induced gap opening. In order to incorporate electronic correlation effect and track bandgap variation over $Q$, we add the nearest-neighbor intersite Fock term \cite{Huang14PhysRevB.90.195114,Jeckelmann02PhysRevLett.89.236401,Campo10JPCM055602}
\begin{equation}\label{eq:Hubb}
    H_{V}=V\sum_{\langle i,j\rangle,\sigma\sigma'}n_{i\sigma}n_{j\sigma'},
\end{equation}
where $n_{i\sigma}=c_{i\sigma}^{\dagger}c_{i\sigma}$ is the spin $\sigma$-resolved occupation number at site-$i$. The main effect of this term is that it renormalizes the hopping integral between intersite bond and adds many body correlation effects (see Supplemental Material, SM \cite{supp} for details). Taking typical parameter values of $t=1$, $\Delta=0.2$, and $V=0.4$, we solve this model self-consistently in the mean-field scheme. We would like to emphasize that the Hubbard term is not included here, as it does not affect the main physical picture. The SOC term $\lambda \tau_z\sigma_z$ is also not critical for such a phase transition, which is omitted as well. Note that while such a model was used to analyze topological phase transition, charge density wave, and ionic chiral displacements \cite{Taddia17PhysRevLett.118.230402,Chen25PRLbfll-sdrb,Parida25kxqp-57vj}, we suggest that it could also describe the NP--FE phase control that has not been investigated previously. Although being a simplified 1D model, this minimal Hamiltonian effectively describe how the near-gap light reshapes the structural energy landscape and captures the essential physics of the DFT results in realistic materials (see SM \cite{supp} for a detailed comparison). In detail, we plot the band dispersion at $Q=0$ at the half-filling (black curve) in Fig. \ref{fig:tb}(a). It opens a direct bandgap of $E_g=0.17t$ at the $\pm X$ point. If we artificially apply a sizable e-h pairs (for instance, $n_{eh}=0.2~e/\rm f.u.$) in the valence and conduction bands, we clearly see that the bandgap increases to $0.20t$ (green curve), yielding an enhancement of $\Delta E_g=0.03t$. This indicates a clear deviation from the single particle rigid-band approximation, well-consistent with our DFT results (shown below). In order to explore the physical origin of such an enhancement, we individually tune the correlation parameter $V$ and e-h concentration $n_{eh}$ while keeping other parameters unchanged. The simulation results are depicted in Fig. \ref{fig:tb}(b). Under finite excitation, the gap enhancement $\Delta E_g$ monotonically increases with $V$, indicating the significance of intersite correlation. This trend is well-consistent with our DFT results below, confirming that the single-particle model is not sufficient to quantify the potential energy variation under light excitation.

Furthermore, we adjust the parameter $Q$ between $[-0.02,0.02]a$ and calculate their band variation. As $|Q|$ increases, $E_g^{Q}$ monotonously enhances, also governed by $V$ that dresses effective hopping integral \cite{supp}. The optical susceptibility function is computed as
\begin{equation}\label{eq:dielect}
    \chi(\omega)=\frac{e^2}{\varepsilon_0}\int [dk]\sum_{c,v}\frac{|\langle u_{vk}|\nabla_k|u_{ck}\rangle|^2}{E_{ck}-E_{vk}-\hbar\omega-i\Gamma}.
\end{equation}
Here, $\varepsilon_0$ is vacuum permittivity, and $c$ ($v$) represents conduction (valence) band index. The integral is performed in $D$-dimensional first Brillouin zone, with $\int[dk]=\int d^Dk/(2\pi)^D$. We take a universal broadening factor $\Gamma=0.025t$ to phenomenologically incorporate disorder, impurity, and thermal effect. The real part $\chi'(\omega)$ [Fig. \ref{fig:tb}(c)] and imaginary part $\chi''(\omega)$ [Fig. \ref{fig:tb}(d)] exhibits clear Kramers-Kronig relationship.

When the incident photon energy ($\omega_1$) is slightly below the absorption threshold, its effect on phase stability can be described by the off-resonant impulsive Raman scattering \cite{Zhou21advs.202003832,Zhou25JPCL.4c03047,Shukla26NMs41563-025-02433-1,Sun24PhysRevB.110.104301,Huang26PRL71fd-m6rj}. Physically, it applies a ponderomotive potential in the Floquet framework, reducing the total potential landscape via
\begin{equation}\label{eq:sc}
    \Phi_{\mathrm{sc}}=-\frac{1}{4}\varepsilon_0\vec{E}_{\omega_1}\cdot\tensor{\chi}'(\omega_1,Q)\cdot\vec{E}_{-\omega_1}.
\end{equation}
Here, $\vec{E}_{\omega_1}$ is the alternating electric field vector. Such an effect has been experimentally observed very recently in different systems such as $\mathrm{LiNbO}_3$, $\mathrm{MoTe}_2$, and 2D perovskites \cite{Shi25adma.202510524,Shi23NCs41467-023-41291-w,Zhang25acsnano.5c10861,Shukla26NMs41563-025-02433-1}, but its effect on FE-NP transition not yet explored. In order to see the trend clearly, we choose a typical $\hbar\omega_1=0.1t$ and plot the $\chi'(\omega_1,Q)$ [Fig. \ref{fig:tb}(e)]. One observes that the $Q=0$ states possesses the largest $\chi'(\omega_1)$. It decays exponentially under finite $|Q|$. From Eq. (\ref{eq:sc}), the energy change is proportional to light intensity $I_0=\frac{1}{2}c_0\varepsilon_0|\vec{E}|^2$ ($c_0$ being the speed of light in vacuum) and reduces by $\chi'(\omega_1,Q)$. Hence, the off-resonant light tends to stabilize the centrosymmetric NP state. This process can operate without direct excitation of infrared-active phonons (and trigger Raman-active modes via nonlinear phononic interactions) or generating e-h pairs, thus is distinct from conventional electron-phonon coupling scheme. We perform a phenomenological analysis according to the Landau theory (see SM for details \cite{supp}), implying that such a phase transition follows the continuous nature without involving latent heat, and the critical light fluence for such a phase transition is proportional to $E_g^{Q=0}$. It would occur collectively with an ultrafast kinetics, while the nucleation-growth process can be avoided.

As the photon energy approaches near $E_g$, photon absorption occurs as $\chi''(\omega)$ becomes finite. Choosing a typical near-resonant value of $\omega_2=0.2t$, the $\chi''(\omega_2,Q)$ also decays with $|Q|$ [Fig. \ref{fig:tb}(e)]. Since photoexcitation generally enhances the total energy \cite{Yang09PhysRevLett.103.186802,Qiu13PhysRevLett.111.216805}, this implies that under near-resonant light the NP state would gain more energy than larger $|Q|$. Such a structure-dependent steady-state $n_{eh}$ under continuous illumiation is balanced by photon e-h excitation and recombination through leakage and depolariation. Therefore, one could expect a light-induced FE state. From the model, we estimate the total energy in the mean field regime and find that the total potential $\Phi(Q)$ shows a double well profile, clearly indicating a NP-FE transition (see SM \cite{supp} for details). According to phenomenological analysis using Landau theory \cite{supp}, in contrast to the off-resonant case, this NP-FE transition exhibits a first-order nature. The critical light fluence depends on $E_g^{Q=0}$ and how it varies with displacement $Q$, which is proportional to the deformation potential strength \cite{Puri26PRL3z14-zzj7}.

\textit{Realistic material examples.}---Armed with this simple yet general model, we next perform DFT calculations (computational details in SM \cite{supp}) on realistic materials to quantify the light frequency and intensity for the phase transitions. Here, we scrutinize the family of group IV($\mathrm{Ge},\mathrm{Sn},\mathrm{Pb}$)--VI($\mathrm{S},\mathrm{Se},\mathrm{Te}$) binary compounds and consider their potential phase transition in (distorted) cubic phase. According to previous works \cite{BarrazaLopez2021,Wiedemeier1978,Littlewood1980a,Littlewood1980b,thermalexpansion,PhysRevB331492,NPGAsiaMaterials,chattopadhyay,POLYMORPHISM,badachhape1962,ONODERA1986240,Strained,alexandrov1987,PhysRevLett.25.159} and our calculations, in general, the telluride compounds are more favored for showcasing our prediction than the selenides and sulfides. The rocksalt phase ($Fm\bar{3}m$) may transit into a distorted structure with the cation and anion dimerize along the $[111]$ direction ($R3m$). They are also well-known phase change materials that are widely studied. Hence, we use tellurides as the primary material platforms, while calculations for the remaining IV--VI compounds are presented in the SM \cite{supp} as constrained-structure references. They all exhibit direct bandgap semiconducting throughout the phase transition path.

Without e-h excitation the $\mathrm{GeTe}$ exhibit spontaneous FE order, while the $\mathrm{SnTe}$ and $\mathrm{PbTe}$ could show rocksalt NP phase. We tabulate their basic lattice constant and geometric structure in SM \cite{supp}, which agree well with previous computational and experimental works \cite{Boschker17C7CE01040K,Zhang09PhysRevB.80.024304}.

\begin{figure}[tb]
    \centering
    \includegraphics[width=0.48\textwidth]{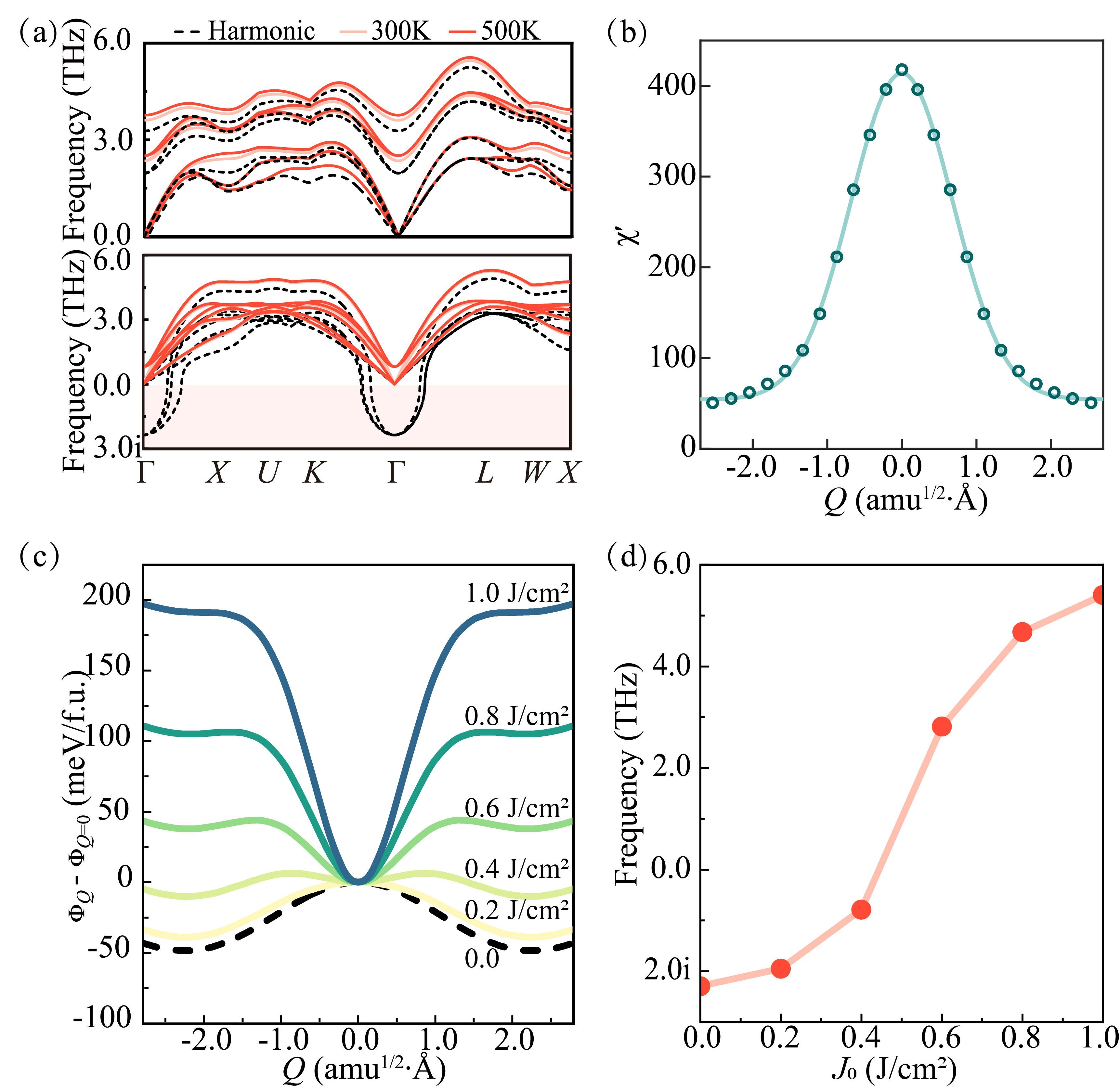}
    \caption{DFT results for off-resonant light induced FE-NP transition of $\mathrm{GeTe}$. (a) Harmonic (dashed curve) and anharmonicity-dressed (solid curves) phonon dispersions for (top panel) $R3m$ and (bottom panel) $Fm\bar{3}m$ phases. The LO-TO (longitudinal and transverse optical) splitting is not included here, which does not affect the main results. (b) $\chi'(\omega_1)$ for different $Q$ along the $[111]$ direction. The variation of (c) light-induced potential $\Phi=\Phi_0+\Phi_{\mathrm{sc}}$ and (d) typical harmonic frequency of $Fm\bar{3}m$ at $\Gamma$, under different light fluence $J_0$. Here, we assume a typical irradiation time of $5\,\mathrm{ps}$ onto the sample.}
    \label{fig:GeTe}
\end{figure}

\textit{FE to NP transition under off-resonant light.}---We take the $\mathrm{GeTe}$ as an exemplary material to evaluate the off-resonant light effect. According to calculations, the FE phase ($Q=2.3\,\mathrm{amu}^{1/2}\cdot\angstrom$) is energetically lower than the NP phase by $48.5\,\mathrm{meV/f.u.}$. We calculate the phonon spectrum to further investigate their dynamical stability. According to harmonic oscillation approximation, the phonon dispersion of both $R3m$ and $Fm\bar{3}m$ phases are plotted as dashed curves in Fig. \ref{fig:GeTe}(a). One sees that the former is dynamically stable, while the latter exhibits a pronounced imaginary frequency of $2.3i\,\mathrm{THz}$ at $\Gamma$. This corresponds to a FE mode with irreducible representation of $T_{1u}$. We then include the anharmonic corrections using self-consistent phonon calculation approach \cite{Tadano15PhysRevB.92.054301,Zhou14PhysRevLett.113.185501} to evaluate the phonon spectrum under finite temperature, indicating its feasibility under thermal condition. The electronic band dispersion can be seen in SM \cite{supp}, which gives $E_g^{Q=2.29}=0.61\,\mathrm{eV}$ and $E_g^{Q=0}=0.08\,\mathrm{eV}$. The former is much larger than the latter, consistent with the results of the TB model.

We next track the $\chi'(\omega_1)$ at a chosen $\hbar\omega_1=0.03\,\mathrm{eV}$, lower than $E_g^{Q=0}$, to ensure no direct photon absorbance. Figure \ref{fig:GeTe}(b) plots its variation over $Q$. It exhibits a symmetric pattern, with $\chi^{\prime}(\omega_1,Q=0)=418$ giving the largest value. As $|Q|$ increases, $\chi'(\omega_1)$ exponentially reduces. According to Eq. (\ref{eq:sc}), the ponderomotive potential at $Q=0$ drops faster than finite $Q$, scaled by light radiation fluence $J_0$. We plot the potential profile (relative to $Q=0$) $\Phi(\omega_1,Q)=\Phi_0(Q)+\Phi_{\mathrm{sc}}(\omega_1,Q)$ in Fig. \ref{fig:GeTe}(c). One sees that as $J_0$ increases, the original double well potential transforms into a single well profile, giving an off-resonant light induced FE to NP transition. Furthermore, we fit the $\Gamma$-centered harmonic infrared mode $T_{1u}$ phonon frequency variation under $J_0$ [Fig. \ref{fig:GeTe}(d)], which becomes real and positive with increasing $J_0$ of $\sim 0.4~\mathrm{J/cm^2}$. This could serve as a fingerprint character that can be measured in experiments. In addition to the time-resolved synchrotron diffraction \cite{Fons10PhysRevB.82.155209} and optical absorption \cite{Marini21PhysRevLett.127.257401}, we propose another potential noncontact optical detection scheme for the symmetry change between FE and NP phase, the second harmonic generation (SHG). As a second order nonlinear optical process, it is sensitive to centrosymmetry. We calculate the SHG susceptibility function along the polar axis, reaching a value of $5.6 \,\mathrm{nm/V}$ \cite{supp}. This is sufficiently large to be measured experimentally.

\textit{Symmetry reduction under $Q$-dependent photoexcitation.} We then move to $\mathrm{Sn}X$ and $\mathrm{Pb}X$ that can be stabilized at the centrosymmetric NP phase. In the main text, we take $\mathrm{SnTe}$ as an example to conduct the calculations, and similar results of other systems are reported in SM \cite{supp}. At $Q=0$, it shows a direct bandgap of $E_g^{Q=0}=0.25\,\mathrm{eV}$ at the BZ corner $L=(1/2,1/2,1/2)$. It increases to $0.58\,\mathrm{eV}$ with a displacement $Q=2.0\,\mathrm{amu}^{1/2}\cdot\angstrom$ \cite{supp}. Such a large variation of $E_g$ implies a strong contrast in the absorbance of different $Q$ under a near-gap light illumination. In realistic device geometries, the electrodes and heterointerfaces create carrier-extraction paths to shorten the carrier lifetime, especially for low gap semiconductors. Hence, at the quasi-steady state, the photoexcited carrier density can be described as $Q$-dependent between generation and recombination, $n_{eh}(Q)$, under a fixed light fluence.

\begin{figure}[bt]
    \centering
    \includegraphics[width=0.48\textwidth]{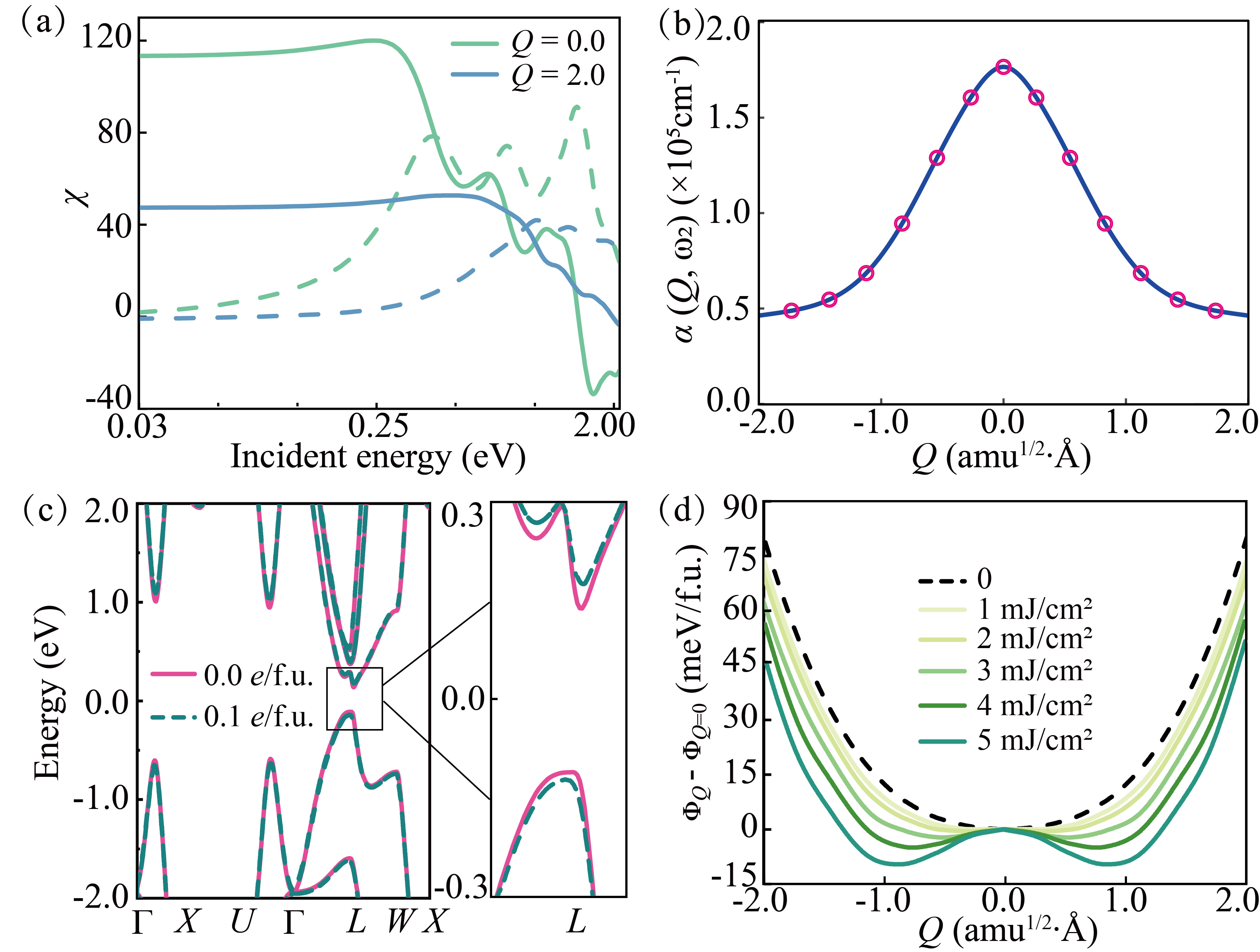}
    \caption{Constrained DFT results for $\mathrm{SnTe}$ phase transition. (a) Frequency-dependent electronic susceptibility functions for $Q=0.0$ and $Q=2.0\,\mathrm{amu^{1/2}\cdot\angstrom}$. (b) $Q$-dependent absorbance coefficient $\alpha(\omega_2,Q)$ at $\hbar\omega_2=0.4\,\mathrm{eV}$, slightly above $E_g^{Q=0}$. (c) Band dispersion along the high symmetric $k$-path for $Fm\bar{3}m$ phase with $n_{eh}=0$ and $n_{eh}=0.1\,e/\mathrm{f.u.}$. (d) Light-induced potential $\Phi_{\mathrm{tot}}(Q)=\Phi_0+\Phi_{\mathrm{abs}}(n_{eh})+\Phi_{\mathrm{sc}}$, relative to $Q=0$, under different light fluence $J_0$. Here, the incident fluence of $2.25\,\rm mJ/cm^2$ corresponds to the photoexcited carrier density $n_{eh}=0.1\,e/\rm f.u.$ at $Q=0$ state, which reduces for larger $|Q|$, see Ref. \cite{supp} for details. Hence, the maximum photoexcited concentration here is below $0.25~e/\rm f.u.$ under $\hbar\omega_2=0.4\,\mathrm{eV}$ irradiation.}
    \label{fig:SnTe}
\end{figure}

In Fig. \ref{fig:SnTe}(a), we plot the frequency-dependent susceptibility functions for $Q=0$ and a typical $R3m$ phase of $Q=2.0\,\mathrm{amu^{1/2}\cdot\angstrom}$. The smearing factor is taken to be $0.10\,\mathrm{eV}$. Excitonic effects are expected to be weak due to strong dielectric screening in bulk phase and small gap values. The absorbance coefficient $\alpha(\omega_2,Q)=2\omega_2\mathrm{Im}[\chi'(\omega_2,Q)+i\chi''(\omega_2,Q)]^{1/2}/c_0$ at a selected photon energy $\hbar\omega_2=0.4\,\mathrm{eV}$ (wavelength of $\sim 3.1\,\mathrm{\mu m}$) is computed. It is evident that the NP state has the largest $\alpha=1.7\times 10^5\,\mathrm{cm}^{-1}$, which quickly reduces when $|Q|$ increases [Fig. \ref{fig:SnTe}(b)]. To estimate the total light-excited $n_{eh}$, we adopt a thin film sample with its thickness of $d=200\,\mathrm{nm}$, then one has $n_{eh}(\omega_2,Q)=\frac{J_0A(\omega_2,Q)\Omega_{\rm f.u.}}{\hbar\omega_2d}$ with $A(\omega_2,Q)=1-e^{-\alpha(\omega_2,Q)d}$ being the total absorption fraction \cite{Sokolowski95PhysRevB.51.14186} and $\Omega_{\rm f.u.}$ the volume per f.u. under normal incidence. Under a moderate light fluence, the excited carrier concentration is on the order of $0.01-0.1\,e/\mathrm{f.u.}$ \cite{supp}, being a small number and less susceptible for overheating and lattice damage. Here, we assume small scattering or reflection and the steady-state balance between absorption and carrier decay, which does not affect the main $Q$ trend and the order estimate. As such, we expect an enhanced excitation energy $\Phi_{\mathrm{abs}}(Q)$ for smaller $|Q|$.

Upon finite e-h pairs excited near the band edge, the electronic band dispersion varies and total energy enhances from the intact ground state. In order to explicitly see this, we perform occupation constrained DFT (cDFT) calculations with the electron and hole carriers following Fermi-Dirac distribution (with respect to their own quasi-chemical potentials). This strategy has been well-adopted in previous works \cite{Paillard19PhysRevLett.123.087601,Peng24PhysRevLett.132.116601,Furci24PhysRevLett.132.236101,Rurali24PhysRevLett.133.116401,Marini21PhysRevLett.127.257401,Marini22PhysRevB.105.L220406,Pal25NCs41467-025-58592-x,Paillard17PhysRevB.96.045205,Yang24PhysRevB.109.184111,Wu26sciadv.adv9616,Paillard16PhysRevLett.116.247401,Marini21PhysRevB.104.144103}. We cross-check the results calculated from the VASP \cite{Kresse93PRB,Kresse96PRB} and Quantum-ESPRESSO \cite{Giannozzi_2009QE,Giannozzi_2017QE} codes, yielding consistent and almost the same results \cite{supp}. The electronic distribution temperature does not affect the main physics here, see SM \cite{supp}. As a typical example, we depict the excited band structure with $n_{eh}=0.1\,e/\mathrm{f.u.}$ for the NP state in Fig. \ref{fig:SnTe}(c). One sees that its bandgap slightly increases (by $16\,\mathrm{meV}$) compared with that of the ground state, indicating a finite deviation from the rigid-band approximation. This gap enhancement value is comparable to the phase transition energy scale, hence such a correlation effect cannot be simply disregarded.

With the total energy carefully computed for each structural state under their respective $n_{eh}(Q)$, we plot $\Phi_{\mathrm{tot}}(\omega_2,Q)$ under $J_0$ in Fig. \ref{fig:SnTe}(d). The results are consistent with model simulations. One sees that when $J_0$ enhances to $4\,\mathrm{mJ/cm^2}$, $\Phi_{\mathrm{tot}}(\omega_2,Q)$ shows a clear evidence of double well profile with the FE state appears at $Q=\pm 0.83\,\mathrm{amu^{1/2}}\cdot\angstrom$. They are energetically lower than the NP state by $6.1\,\mathrm{meV/f.u.}$. This is significant enough for the FE emergence. Therefore, one can achieve a symmetry reduction phase transition when the photon energy is adjusted close to the NP bandgap, with tunable light absorbance along the phase transition pathway. Our calculated SHG signal for the FE phase reaches $30.4\,\mathrm{nm/V}$ \cite{supp}, sufficiently large for experimental detection. The pump and probe light beams can be chosen in different frequencies, so that no interference problems exist.

\textit{Discussion and conclusion.}---In order to stabilize and engineer the transient light-induced FE phase in a nonvolatile regime, one can apply mechanical strain to further modulate the energetic landscape \cite{Li21NRMs41578-021-00304-0,Cao24Natures41586-024-08099-0}. For instance, we find that the FE phase could exist if a $3\%$ triaxial tensile strain is applied \cite{supp}. Such a strain effect would furthermore aid the phase transition kinetics, adding an additional term in the total potential expression. Note that photoexcitation also leads to finite strain field, yielding photostriction in various systems \cite{Peng2022,Puri26PRL3z14-zzj7,Zhou20PRR2.022059}. Although our work focuses on group-IV tellurides as examples, the same mechanism is applicable to other polar semiconductors with $Q$-dependent optical susceptibility, such as perovskites $\mathrm{CsGe}X_3\,(X=\rm Cl,Br,I)$, $\rm LiNbO_3$, and $\rm LiTaO_3$.  As the below-gap light could penetrate deeply into the film without significant absorbance at the surface, the phase transition would occur in a coherent and collective way. One can expect a complete reversible light-stabilized phase transition with negligible residual back-stress in a thin film sample. 

In summary, we adopt a minimal TB model with generalized correlations and \textit{ab initio} simulations to unravel how near-gap light could reshape the total adiabatic potential landscape and trigger phase transition between FE and NP states, with a frequency-tunable nonvolatile structural control. We further predict that group-IV tellurides with small direct bandgaps are optimal testbeds, in which phase transition and hidden phase could emerge under moderate mid-infrared light. Our work extends the light-induced phase transition mechanisms with constant e-h pairs under well-above gap light and closed boundary condition, suggesting a contactless strategy for ferroic order emergence that are not thermally accessible in the usual pressure-temperature phase diagram.

\begin{acknowledgments}
\textit{Acknowledgments.}---J.Z. acknowledges valuable discussions with Z. Jiang, P. Chen, and H. J. Zhao on the phase control mechanisms and simulation details. This work is supported by the National Natural Science Foundation of China (NSFC) under Grant No. 12374065. The HPC platform of Xi'an Jiaotong University is also acknowledged.
\end{acknowledgments}

\textit{Data availability.}---The data that support the findings of this article are publicly available \cite{ScienceDB}.


\providecommand{\noopsort}[1]{}\providecommand{\singleletter}[1]{#1}%

\end{document}